\documentclass[twocolumn,aps,prl,superscriptaddress]{revtex4-1}
\usepackage{graphicx}
\usepackage{dcolumn}
\usepackage{bm}
\newcolumntype{w}[1]{D{.}{.}{#1}}
\newcolumntype{.}{D{x}{}{-1}}

\begin{document}

\title{Frequency metrology of helium around 1083 nm and determination of the
  nuclear charge radius}

\author{P. Cancio Pastor}
   \email{pablo.canciopastor@ino.it}
%   \homepage{http://www.ino.it, http://lens.unifi.it}
\author{L. Consolino}
\author{G. Giusfredi}
\author{P. De Natale}
\affiliation{Istituto Nazionale di Ottica-CNR (INO-CNR), Via Nello Carrara 1, I-50019 Sesto Fiorentino, Italy}
\affiliation{ European Laboratory for Non-Linear Spectroscopy (LENS) and
  Dipartimento di Fisica,
Universit\'a di Firenze\\ Via Nello Carrara 1, I-50019 Sesto Fiorentino, Italy}
\author{M. Inguscio}
\affiliation{ European Laboratory for Non-Linear Spectroscopy (LENS) and
  Dipartimento di Fisica,
Universit\'a di Firenze\\ Via Nello Carrara 1, I-50019 Sesto Fiorentino, Italy}
\author{V. A. Yerokhin}
\affiliation{St. Petersburg State Polytechnical
  University,
Polytekhnicheskaya 29, St. Petersburg 195251, Russia}
\author{K. Pachucki}
\affiliation{Faculty of Physics, University of Warsaw, Hoza 69, 00-681 Warsaw, Poland}

\begin{abstract}

We measure the absolute frequency of seven out of the nine allowed transitions
between the 2$^3${\it S} and 2$^3${\it P} hyperfine manifolds in a metastable $^3$He beam
by using an optical frequency comb synthesizer-assisted spectrometer.
The relative uncertainty of our measurements ranges from
$1\times 10^{-11}$ to $5\times 10^{-12}$, which is, to our knowledge,
the most precise result for any optical $^3$He transition to date.
The resulting $2^3${\it P}-2$^3${\it S} centroid frequency is $276\,702\,827\,204.8\,(2.4)$~kHz. 
Comparing this value with the known result for the $^4$He centroid and
performing {\em ab initio} QED calculations of the $^4$He-$^3$He isotope
shift, we extract the difference of the squared nuclear 
charge radii $\delta r^2$ of $^3$He and $^4$He. Our result for
$\delta r^2=1.074 (3)$ fm$^2$ disagrees by about $4\,\sigma$ 
with the recent determination [R.~van Rooij {\em et al.}, Science {\bf 333}, 196  (2011)]. 
\end{abstract}

\maketitle

Spectacular progress of experimental techniques, achieved in the last decades, has
made precision spectroscopy of light atoms a unique tool for the determination of
fundamental constants and properties of atomic nuclei.
The underlying theory, quantum electrodynamics
(QED), allows one to calculate atomic properties {\em ab initio} and keep
control of the magnitude of uncalculated effects. Possible discrepancies
between theory and experiment may signal a lack of our knowledge of
details of the interactions between electrons, nuclei, and other particles.
An important recent example is the discrepancy of the proton charge radius
derived from the spectroscopy of the electronic and muonic hydrogen
\cite{pohl:10}. This discrepancy is still unresolved and
might lead to important consequences, such as
a change of the accepted value for the Rydberg constant (which was, up to now,
considered to be one of the best known fundamental constants) or discovery
of unknown effects in the electromagnetic lepton-nucleus interaction.

Another important disagreement reported recently
concerns the charge radii of helium isotopes. Specifically, the difference of
the squares of the charge radii of the $^3$He and $^4$He nuclei determined from the
2$^1${\it S}-2$^3${\it S} transition \cite{rooij:11} was reported to differ
by about 4 standard deviations ($\sigma$) from that derived using the
2$^3${\it P}$_0$-2$^3${\it S} transition \cite{shiner:95}. In this Letter, we aim to
resolve this discrepancy by performing an independent, high-precision
measurement of the $^4$He-$^3$He isotope shift, on the one hand, and by
advancing the theory of the helium isotope shift, on the other hand.

In the experimental part, we measure the absolute frequency of seven out of the
nine allowed transitions between the 2$^3${\it S} and 2$^3${\it P} hyperfine manifolds of
$^3$He with a precision ranging from $1\times 10^{-11}$ to $5\times
10^{-12}$. To the best of our knowledge, these are currently the most
accurate measurements for any optical $^3$He transition. In the theoretical part,
we perform a rigorous QED calculation of the isotope shift of the centroid of
the 2$^3${\it P}-2$^3${\it S} transition, identifying several corrections omitted in the
previous studies and carefully examining the uncertainty due to uncalculated
effects. Combining the experimental and theoretical results, we obtain the
difference of the squares of the charge radii of the $^3$He and $^4$He
nuclei. The improved theory is also applied for reexamination of the
previous experimental result \cite{rooij:11}.

\begin{figure}[h]
\includegraphics[width=8.6 cm]{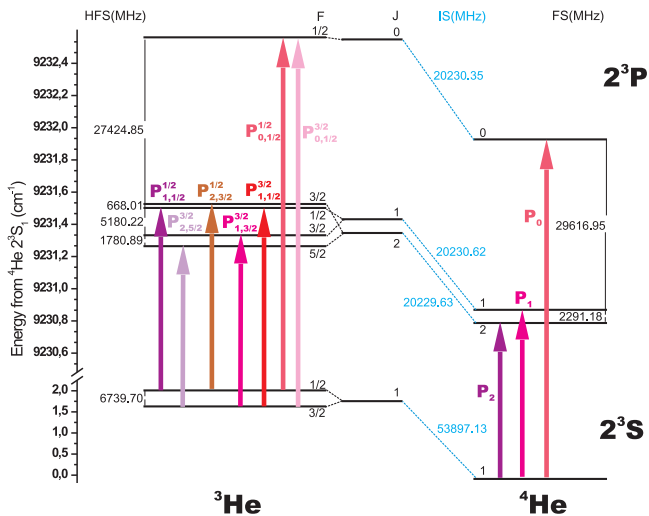}
\caption{\label{figura1} (color online) Level scheme of the 2$^3${\it S} and 2$^3${\it P} manifolds of
         $^3$He and $^4$He isotopes. $^3$He HFS,  $^4$He fine structure (FS) and $^4$He -$^3$He IS splittings are shown.}
\end{figure}

The structure of the 2$^3${\it S} and 2$^3${\it P} levels in $^3$He and $^4$He is shown
in Fig.~\ref{figura1}. In the following, we will use the shorthand notation
{\it P}$^{F_S}_{J_P,F_P}$ to denote the
2$^3${\it S}$_{1,F_S}\rightarrow$2$^3${\it P}$_{J_P,F_P}$ hyperfine transition in $^3$He
and {\it P}$_{J_P}$ to denote the 2$^3${\it S}$_{1}\rightarrow$2$^3${\it P}$_{J_P}$ transition
in $^4$He. We now address the measurement procedure applied to the
seven transitions between the 2$^3${\it S} and 2$^3${\it P} hyperfine
manifolds of $^3$He. The remaining two allowed transitions, {\it P}$^{1/2}_{1,3/2}$ and
{\it P}$^{3/2}_{2,3/2}$, have a very weak intensity due to the
hyperfine suppression~\cite{sulai:08,consolino:}, which prohibits their
measurement with our spectroscopic setup.

Multiresonant precision spectroscopy was performed by using
the optical frequency comb synthesizer (OFCS)-assisted laser system
at 1083 nm described in Ref.~\cite{consolino:11}.
In each run, quasisimultaneous saturation spectra  of two out of these seven
hyperfine transitions were recorded in an absolute frequency scale,
and the frequency center was measured by a fitting procedure~\cite{consolino:11}.
In this way, absolute frequencies of each transition and their differences
are measured minimizing possible time-dependent systematic errors.
This procedure is repeated for about 200 runs for each transition.
During these runs, different transitions 
were coupled to each other,
and for each couple the multiresonant probe laser system was interchanged,
in order to randomize, as much as possible, the measurement procedure.
The final results, together with the error budget, are summarized in
Table~\ref{tabella1}.

\begin{figure}[h]
\includegraphics[width=8.6 cm]{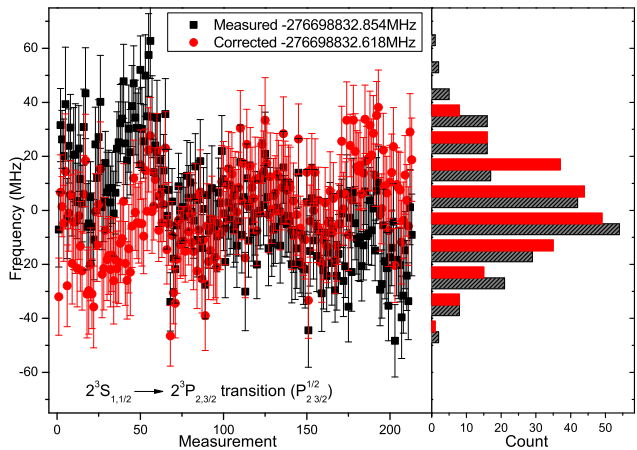}
\caption{\label{figura2} (color online) Day-by-day correction of the RS+LS+2DS shift for the measurements of the {\it P}$^{1/2}_{2,3/2}$ hyperfine transition. Left graph: the squares (black) are the measured data, and the circles (red) are the corrected data. A different mean frequency was subtracted to bring data in the same frequency vertical scale for a clear comparison. Right graph: bar graph distribution of the measured (dashed black bars) and corrected (fill red bars) frequencies.}
\end{figure}

\begingroup
\squeezetable
\begin{table}[floatfix]
\caption{\label{tabella1}Absolute frequency measurements of
{\it P}$^{F_S}_{J_P,F_P}$ $^3$He hyperfine transitions:
statistical results and systematic error budget, in kHz.
Uncertainties are given in parentheses.
}
\begin{ruledtabular}
\begin{tabular}{lccc}
Transition
&Statistical value\footnote{Each measurement was corrected by the day-by-day RS+LS+2DS shifts (see text for details).}&Zeeman\footnote{Because of residual magnetic fields
($<0.03\,\mu$T).}&Final\footnote{The OFCS error of 10$^{-12}$$\times$statistical
value (kHz) and the 1DS error of 0.8~kHz were added in quadrature
in the final uncertainty.} \\
 &&shift&result  \\
\hline
\hline
{\it P}$_{1,1/2}^{1/2}$&276\,698\,164\,610.4(1.8)&(0.233)
&276\,698\,164\,610.4(2.0) \\
{\it P}$_{2,5/2}^{3/2}$&276\,698\,611\,209.1(3.1)&(0.148)
&276\,698\,611\,209.1(3.2)\\
{\it P}$_{2,3/2}^{1/2}$&276\,698\,832\,617.9(2.5)&(0.303)
&276\,698\,832\,617.9(2.5)\\
{\it P}$_{1,3/2}^{3/2}$&276\,700\,392\,099.8(0.9)&(0.369)&
27 6\,700\,392\,099.8(1.3)\\
{\it P}$_{1,1/2}^{3/2}$&276\,704\,904\,311.7(2.3)&(0.233)
&276\,704\,904\,311.7(2.4)\\
{\it P}$_{0,1/2}^{1/2}$&276\,726\,257\,468.9(1.1)&(0.932)
&276\,726\,257\,468.9(1.7)\\
{\it P}$_{0,1/2}^{3/2}$&276\,732\,997\,170.4(2.3)&(0.466)
&276\,732\,997\,170.4(2.5)
\\
\end{tabular}
\end{ruledtabular}
\end{table}
\endgroup
Particular attention was paid to single out and quantify possible systematic
errors. As in our previous $^4$He measurements~\cite{cancio:04, giusfredi:05},
the main systematic correction was due to recoil-induced mechanical shift
(RS)~\cite{minardi:99:1}. As in that case, we calculate its contribution for
each transition by solving the atomic Bloch equations, including  ac Stark
shift [light shift LS], and taking into account the atom dynamics during
interaction with the laser, in the present experimental
conditions. Second-order Doppler (2DS) correction due to the
longitudinal velocity distribution of the atoms in the $^3$He metastable beam
was also included in this calculation. Since RS is like an accumulated recoil
during laser-atom interaction time, it is strongly dependent on the metastable
atomic flux in our experimental setup, and hence on the dc discharge
conditions used to metastabilize the $^3$He atomic beam. We have noticed that
such conditions changed during a day of measurements, due to progressive
saturation of the filtering system for contaminant gases, used in the $^3$He
gas recycling line inserted in the atomic beam. In fact, we monitored this
change by measuring the atomic longitudinal velocity distribution behavior
during a day. From this data, an averaged longitudinal velocity distribution
for each day is determined, which enters as a parameter in the
RS+LS+2DS calculation. All measurements for each transition are
corrected by the corresponding day shift. The final frequency is calculated as
a statistical average of all corrected measurements performed for each
transition. Such a procedure is shown in Fig.~\ref{figura2} for the
{\it P}$^{1/2}_{2,3/2}$ transition, where about 180 measurements without
(squares) and with (circles) RS+LS+2DS day-shift corrections
are reported.  As a result, a Gaussian distribution of the corrected
measurements is shown in the fill-bar graph of Fig.\ref{figura2}, witnessing
that our data have \lq{}\lq{}white" statistical fluctuations.

Systematics uncertainties due to first-order Doppler shift (1DS),  OFCS
accuracy, and Zeeman shift, have been added in quadrature to the
statistical one, as shown in Table~\ref{tabella1}.  1DS was avoided due
to the saturation spectroscopy configuration~\cite{cancio:04}, but with an
error of 0.8 kHz, due to the achievable angular accuracy between  the forward
and the backward interacting laser beams. The Rb-GPS disciplined quartz
oscillator, used in our OFCS,  guarantees a relative accuracy of 10$^{-12}$,
considered in the error budget. Finally, a residual magnetic field in the atom-laser interaction region lower than 0.03~$\mu$T, gives a
Zeeman shift uncertainty for each transition (Table~\ref{tabella1}, third column),
without shifting the transition barycenter.
The total accuracy of our measurements, summarized in Table~\ref{tabella1},
ranges from 1$\times$10$^{-11}$ to 5$\times$10$^{-12}$, which is currently the
best reported result for any optical transition in $^3$He.

An independent check for the accuracy of our measurements is made by extracting the
known value 
of the hyperfine splitting (HFS) of the metastable 2$^3${\it S} state from the
transitions in Table~\ref{tabella1}. The two differences
{\it P}$^{1/2}_{0,1/2}$- {\it P}$^{3/2}_{0,1/2}$ and {\it P}$^{1/2}_{1,1/2}$-{\it P}$^{3/2}_{1,1/2}$
yield the values of $6\,739\,701.5\,(3.0)$ and $6\,739\,701.3(3.1)$~kHz,
respectively, which are consistent with each other and in perfect agreement
with the more accurate result of $6739701.177\,(16)$~kHz \cite{rosner:70}.

Comparison of our measurements with the previous experimental results is given
in Table~\ref{tabella2}. The centroid values of the 2$^3${\it P} and 2$^3${\it S} energies are defined
as an average over all fine and hyperfine sublevels,
\begin{eqnarray} \label{centroid}
  E(2^3L) &=&\frac{\sum_{J, F} (2\,F+1)\,E(2^3L_{J,F})}
                {(2\,I+1)\,(2\,S+1)\,(2\,L+1)}\,,
\end{eqnarray}
\begingroup
\begin{table}[floatfix]
\caption{\label{tabella2}
Comparison with prior measurements, in kHz.
}
\begin{ruledtabular}
\begin{tabular}{l.l}
{$^3$He 2$^3${\it P}$-$2$^3${\it S} centroid}  &276\,702\,x827\,204.8\,(2.4)\\
                                 &276\,702\,x827\,145\,(77)\footnote{Evaluated
  by combining the $^4$He {\it P}$_2$ frequency \cite{shiner:94}, the $^4$He-$^3$He
  {\it P}$_2$-{\it P}$_{0,1/2}^{3/2}$ interval \cite{shiner:95}, the $^3$He 2$^3${\it S}
  HFS~\cite{rosner:70}, and the 2$^3${\it P} HFS~\cite{morton:06:cjp}.} \\[0.4em]
{$^4$He 2$^3${\it P}$-$2$^3${\it S} centroid} &276\,736\,x495\,649.5\,(2.1)& \cite{cancio:04}\\
                  &276\,736\,x495\,580\,(70) & \cite{shiner:94}\\[0.4em]
{$^4$He {\it P}$_2-^3$He {\it P}$^{3/2}_{0,1/2}$} &x810\,594.3\,(3.3)\footnote{Evaluated
  by combining the {\it P}$^{3/2}_{0,1/2}$ interval from Table~\ref{tabella1}, the {\it P}$_0$ frequency \cite{cancio:04}, and the 2$^3${\it P}$_0$-2$^3${\it P}$_2$ interval \cite{smiciklas:10}.}\\
                                      &x810\,599\,(3) &\cite{shiner:95} \\
                                      &x810\,608\,(30) &  \cite{zhao:91}\\[0.4em]
{$^3$He {\it P}$^{3/2}_{0,1/2}-^4$He {\it P}$_1$} &1\,x480\,582.1\,(3.2)\\
                                      &1\,x480\,573\,(30) & \cite{zhao:91}
\end{tabular}
\end{ruledtabular}
\end{table}
\endgroup
where 2$^3${\it L} = 2$^3${\it S} and 2$^3${\it P} for {\it L}$ = 0$ and $1$, respectively.
To the best of our knowledge, there are no published measurements of
2$^3${\it P}-2$^3${\it S} HFS frequencies. Therefore, we obtain the ``previous
experimental value'' of the $^3$He 2$^3${\it P}-2$^3${\it S} centroid frequency as a
combination of several experiments and the calculated HFS intervals
(see footnote in Table~\ref{tabella2}).
The previous result is in agreement with our measurement but 33 times
less accurate.
The $^4$He 2$^3${\it P}-2$^3${\it S} centroid was measured by us previously
\cite{cancio:04}, in agreement with the independent determination by Shiner
{\em et al.} \cite{shiner:94}.
In order to check the consistency of our
present measurements on $^3$He with our previous measurements for $^4$He
\cite{cancio:04},  in Table~\ref{tabella2}, a comparison of the
frequency differences of the $^3$He
{\it P}$^{3/2}_{0,1/2}$ and the $^4$He {\it P}$_{1}$ and {\it P}$_2$ intervals with independent
measurements \cite{shiner:95,zhao:91} is reported.

%%%%%%%%%%%%%%%%%%%%%%%%%%%%%%%%%%%%%%%%%%%%%%%%%%%%%%%%%%%%%%%%%%%%%%%
\begingroup
\begin{table}[floatfix]
\caption{
$^4$He--$^3$He isotope shift of the centroid energies,
for the pointlike nucleus, in kHz.
$m_r$ is the reduced mass and $M$ is the nuclear mass.
\label{table:is}
}
\begin{ruledtabular}
  \begin{tabular}{l..}
Contribution & \multicolumn{1}{c}{2$^3$P$-$2$^3$S} & \multicolumn{1}{c}{2$^1$S$-$2$^3$S} \\
\hline\\[-1em]
$m_r\,\alpha^2\,$         &                 12\,412\,458x.1 &  8\,632\,567x.86 \\
$m_r\,\alpha^2\,(m_r/M)$  &                 21\,243\,041x.3 &    -608\,175x.58 \\
$m_r\,\alpha^2\,(m_r/M)^2$&                    13\,874x.6 &         7\,319x.80 \\
$m_r\,\alpha^2\,(m_r/M)^3$&                        4x.6 &               -0x.30 \\
$m_r\,\alpha^4\,$         &                    17\,872x.8 &         8\,954x.22 \\
$m_r\,\alpha^4\,(m_r/M)$  &                   -20\,082x.4 &        -6\,458x.23 \\
$m_r\,\alpha^4\,(m_r/M)^2$&                       -3x.0 &               -1x.84 \\
$m\,\alpha^5\,(m/M)$      &                      -60x.7 &              -56x.61   \\
$m\,\alpha^6\,(m/M)$      &                      -15x.5\,(3.9) &        -2x.75\,(69)  \\
Nuclear polarizability      &                       -1x.1\,(1) &        -0x.20\,(2)  \\
HFS mixing&                       54x.6 &               -80x.69  \\
Total                 &            33\,667\,143x.2\,(3.9)& 8\,034\,065x.69\,(69) \\
Other theory \cite{morton:06:cjp,drake:10}\footnote{Corrected by adding the triplet-singlet
HFS mixing.}  &     33\,667\,146x.2\,(7)&      8\,034\,067x.8\,(1.1)
  \end{tabular}
\end{ruledtabular}
\end{table}
\endgroup
%%%%%%%%%%%%%%%%%%%%%%%%%%%%%%%%%%%%%%%%%%%%%%%%%%%%%%%%%%%%%%%%%%%%%%%

The difference of our results for the 2$^3${\it P}-2$^3${\it S} centroid energy in $^3$He
and $^4$He yields the experimental value of
the isotope shift (IS). Combined with theoretical calculations, the 
experimental IS can be used \cite{marin:94}
to determine the difference of the squared nuclear charge
radii, $\delta r^2 \equiv r^2(^3\mbox{\rm He})-r^2(^4\mbox{\rm He})$.

Numerical results of our calculation of the IS of the
2$^3${\it P}-2$^3${\it S} and 2$^1${\it S}-2$^3${\it S} transitions for the point nucleus
are presented in
Table~\ref{table:is}. As compared to the previous evaluations
\cite{morton:06:cjp,drake:10}, the higher-order recoil
[$m_r\,\alpha^2\,(m_r/M)^3$ and $m_r\,\alpha^4\,(m_r/M)^2$] and the nuclear
polarizability corrections were accounted for and the higher-order QED effects
[$m\,\alpha^6\,(m/M)$] were estimated more carefully. The calculation
extends our previous works \cite{pachucki:09:hefs,pachucki:10:hefs}; its
details will be reported elsewhere. The total uncertainty
comes from the uncalculated higher-order QED effects. Note that
it is much larger than the one reported previously in Ref.~\cite{morton:06:cjp}.

Our definition of the IS differs from that used
previously \cite{shiner:95,rooij:11} by the fact that we
average out not only the hyperfine but also the
fine-structure splitting. The advantage of using
the centroid energy is that theory becomes much more transparent 
and can be directly compared to the experiment.

The difference of the theoretical point-nucleus results in
Table~\ref{table:is} and the experimental IS
comes from the finite nuclear
size (FNS) effect, which can be parametrized as
\begin{eqnarray}
\delta E_{\rm FNS} &=& \frac{2\pi}{3}\,Z\alpha\, m_r^3\,r^2\,
\bigl\langle\delta^{(3)}(r_1)+\delta^{(3)}(r_2)\bigr\rangle\,
 \nonumber \\ && \times
  \biggl[
     1-(Z\alpha)^2\, \ln (Z\alpha\, r) + (Z\alpha)^2f_{\rm rel}\biggr]\,,
\end{eqnarray}
where $f_{\rm rel}$ is the relativistic correction beyond the
leading logarithm. The FNS contribution can be represented as
$\delta E_{\rm FNS} = C\,r^2$, where the coefficient $C$,
according to the above equation, depends very
weakly on $r$. Our calculated values for the coefficient
$C$ are
\begin{eqnarray} \label{C1}
C(2^3{\it P}-2^3{\it S}) &=& -1212.2\,(1)\,\,\, {\rm kHz/fm}^2\,,\\
C(2^1{\it S}-2^3{\it S}) &=& -214.69\,(1)\,\,\, {\rm kHz/fm}^2\,,
\label{C2}
\end{eqnarray}
which can be compared with the previous results
$C(2^3P-2^3S) = -1209.8$~\cite{morton:06}
and $C(2^1S-2^3S) = -214.40$~\cite{drake:05}.

At present, there are three independent measurements of the $^4$He-$^3$He
IS that can be used to infer $\delta r^2$ with a comparable accuracy,
our experiment and those of Refs.~\cite{rooij:11,shiner:95}. Our
theory, summarized in Table~\ref{table:is}, allows us to extract the charge
radius difference $\delta r^2$ consistently from the present experiment
and that of Ref.~\cite{rooij:11}. The results are 
\begin{eqnarray}
\delta r^2(\mbox{\rm this work}) &=& 1.074\,(3)\,{\rm fm}^2\,\label{eqq1},\\
\delta r^2(\mbox{\cite{rooij:11}}, \ \rm reevaluated) &=& 1.028\,(11)\,{\rm fm}^2\,.
\label{eqq2}
\end{eqnarray}
The $\delta r^2$ value of Eq.~(\ref{eqq2}) is by about 1$\sigma$ larger than
that given in Ref.~\cite{rooij:11}, $1.019\,(11)$~fm$^2$, because of the
change in the theoretical IS value.
Using Eq.~(\ref{eqq1}) and the nuclear charge radius of $^4$He \cite{sick:08},
we obtain the root-mean-square radius of the $^3$He isotope to be $1.975\,(4)$~fm. 

The results of Eqs.~(\ref{eqq1}) and (\ref{eqq2}) 
can be also compared with the determination by Shiner {\em et al.},
\begin{eqnarray}
\delta r^2(\mbox{\cite{shiner:95}}) &=& 1.059\,(3)\,{\rm fm}^2\,.
\end{eqnarray}
which used the older isotope shift theory.
We do not reevaluate this result here, since it would 
require improvement in theoretical predictions for HFS intervals.

\begin{figure}
\includegraphics[width=8 cm]{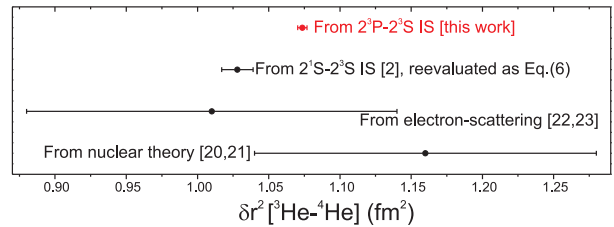}
\caption{\label{figura3} (color online) 
Different determinations of
the difference of the squared nuclear charge radii for  $^3$He
and $^4$He.}
\end{figure}

We observe, as shown in Fig.\ref{figura3}, that the above results for the radius difference $\delta r^2$
are inconsistent with each other. In particular,
a $4\,\sigma$ discrepancy is present between 
our value and that of Ref.~\cite{rooij:11}, for which we do not have a 
satisfactory explanation at present.
Note that both experiments use the same OFCS assisted laser technology,
that most of the theoretical contributions are checked by independent
calculations, and that the determination of the charge radius difference is now performed
consistently within the same theory. The observed discrepancy may be in principle
explained by some hidden systematics in experiments or by
yet unknown effects in the electron-nucleus interaction.

The possibility that some additional effects beyond the standard QED exist
has been discussed in the literature ever since the muonic hydrogen
experiment \cite{pohl:10} raised what is now known as the proton charge
radius puzzle. One of the ways for solving this puzzle is to
investigate similar systems, aiming to confirm or to disprove the
disagreement observed for hydrogen. In the present Letter, we report a 
$4\sigma$ discrepancy for the nuclear charge radius difference of $^3$He 
and $^4$He. 

Precision spectroscopic determination of the nuclear charge
radii of the helium isotopes becomes  today of particular importance, as
the next goal of the muonic hydrogen group from the Paul
Scherrer Institute is the measurement of the Lamb shift in muonic helium 
\cite{antognini:11}. 
This experiment will provide an independent determination of the charge radii of
helium isotopes and will allow one to compare the results obtained by the
spectroscopy of the electronic and muonic atoms, thus hopefully shedding light on the 
proton charge radius problem and on the discrepancy for the helium
charge radius difference.

\end{document}